\documentclass[draft,nomathfonts]{aipproc}

\layoutstyle{6x9}

\def\beq{\begin{equation}}
\def\enq{\end{equation}}
\def\ba{\begin{eqnarray}}
\def\ea{\end{eqnarray}}
\def\bitm{\bibitem}
\def\Mesz{M\'esz\'aros}
\def\siml{\lower4pt \hbox{$\buildrel < \over \sim$}}
\def\simg{\lower4pt \hbox{$\buildrel > \over \sim$}}

\def\s{{\rm s}}
\def\lbr{\linebreak}

\begin{document}

\title{Theories of Early Afterglow}

\author{P. M\'esz\'aros }{
  address={Dept. of Astronomy \& Astrophysics and Dept. of Physics,
Pennsylvania State University,
University Park, PA 16802, USA}
}

\begin{abstract}
The rapid follow-up of gamma-ray burst (GRB) afterglows made possible by
the multi-wavelength satellite {\em Swift}, launched in November 2004,
has put under a microscope the GRB early post-burst behavior, This is
leading to a significant reappraisal and expansion of the standard view of
the GRB early afterglow behavior, and its connection to the prompt
gamma-ray emission. In addition to opening up the previously poorly known
behavior on minutes to hours timescales, two other new pieces in the GRB
puzzle being filled in are the the discovery and follow-up  of short GRB
afterglows, and the opening up of the $z\simg 6$ redshift range.  We review 
some of the current theoretical interpretations of these new phenomena.
\end{abstract}

\maketitle


\section{Challenges posed by new Swift observations}
\label{sec:obs}

Compared to previous satellites, Swift has made a large difference
on two main accounts. First, the sensitivity of the Burst Alert Detector 
(BAT, in the range 20-150 keV) is a factor $\sim$ 5 higher than for the 
corresponding instruments in the predecessor CGRO-BATSE, BeppoSAX and HETE-2. 
Second, Swift can slew in less then 100 seconds in the direction determined 
by the BAT instrument, positioning its much higher angular resolution X-ray 
(XRT) and UV-Optical (UVOT) detectors on the burst \cite{gehr05_thisproc}

As of December 2005, at an average rate of 2 bursts detected per week,
over 100 bursts had been detected by BAT, of which 90\% were followed
promptly with the XRT within $350$ s from the trigger, and about half 
within 100 s \cite{burrows05}, while $\sim 30\%$ were detected with 
the UVOT \cite{roming05}.  Of these, over 23 resulted in redshift 
determinations. Ten short GRB were detected, of which five had detected 
X-ray afterglows, three had optical, and one had a radio afterglow, and
five had a redshift determination.

The new observations brings the total redshift determinations to over 
50 since 1997 when BeppoSAX enabled the first one. The redshifts based 
on Swift have a median $z\simg 2$, which is a factor $\sim 2$ higher 
than the median of those previously culled via BeppoSAX and HETE-2,
\cite{berger05a}. This can be ascribed to the higher sensitivity of BAT 
and the prompt accurate positions from XRT and UVOT, making possible 
ground-based detection at a stage when the afterglow is much brighter.
The highest Swift-enabled redshift so far is in GRB 050904, obtained
with Subaru, $z=6.29$ \cite{kawai05}, the second highest being GRB 050814 
at $z=5.3$, whereas the previous Beppo-SAX era record was $z=4.5$. 
The relative paucity of UVOT detections versus XRT 
detections may be ascribed in part to this higher median redshift, and 
in part to the higher dust extinction at the implied  shorter rest-frame 
wavelenghts for a given observed frequency \cite{roming05}, although 
additional effects may be at work too.

The BAT light curves show that in some of  the bursts which fall in the
``long" category ($t_\gamma \simg 2$ s) faint soft gamma-ray tails 
can be followed  which extend the duration by a factor up to two beyond 
what BATSE could have detected \cite{gehr05_thisproc}.
A rich trove of information on the burst and afterglow physics has come
from detailed XRT light curves, starting on average 100 seconds after 
the trigger, together with the corresponding BAT light curves and spectra.
This suggests a canonical X-ray afterglow \cite{nousek05}
with one or more of the following:\hfill
\lbr
1) an initial steep decay $F_X \propto t^{-\alpha_1}$ with a temporal index
$3 \siml \alpha_1 \siml 5$, and an energy spectrum $F_\nu \propto \nu^{-\beta_1}$
with energy spectral index $1 \siml \beta_1 \siml 2$ (or photon number index
$2 \siml \Gamma=\alpha+1 \siml 3$), extending up to a time $300 \s \siml t_1 
\siml  500 \s$;\hfill
\lbr
2) a flatter decay $F_X \propto t^{-\alpha_2}$ with $0.2 \siml \alpha_2
\siml 0.8$ and energy index $0.7 \siml \beta_2 \siml 1.2$, at times
$10^3 \s \siml t_2 \siml 10^4 \s$;\hfill
\lbr
3) a ``normal" decay $F_X \propto t^{-\alpha_3}$ with $1.1 \siml \alpha_3
\siml 1.7$ and $0.7 \siml \beta_2 \siml 1.2$ (generally unchanged
the previous stage), up to a time $t_3 \sim 10^5 \s$, or in some cases 
longer;\hfill
\lbr
4) In some cases, a steeper decay $F_X \propto
t^{-\alpha_4}$ with $2\siml \alpha_4 \siml 3$, after $t_4\sim 10^5 \s$;\hfill
\lbr
5) In about half the afterglows, one or more X-ray flares are observed,
sometimes starting as early as 100 s after trigger, and sometimes as
late as $10^5 \s$. The energy in these flares ranges from a percent
up to a value comparable to the prompt emission (in GRB 050502b).
The rise and decay times of these flares is unusually steep, depending
on the reference time $t_0$, behaving as $(t-t_0)^{\pm \alpha_{fl}}$ 
with $3 \siml \alpha_{fl} \siml 6$, and energy indices which can be
also steeper than during the smooth decay portions. The flux level after
the flare usually decays to the value extrapolated from the value before
the flare rise.
\begin{figure}[h]
  {\includegraphics[height=.3\textheight,width=.9\textwidth]{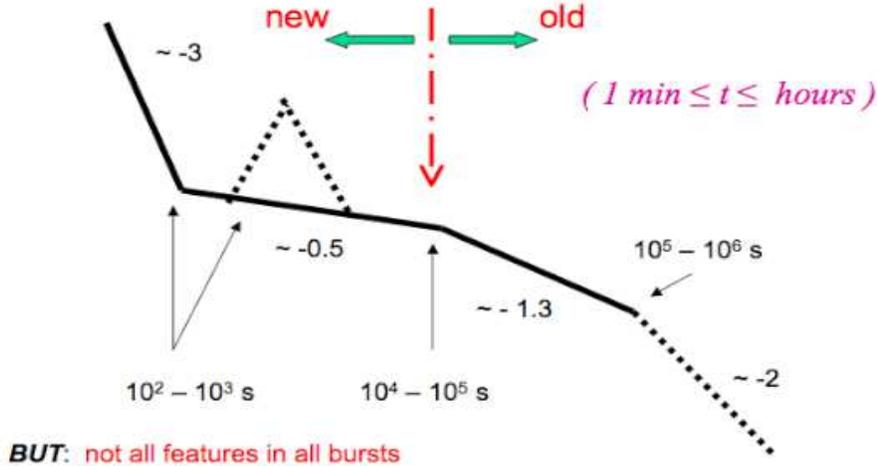}}
\caption{Schematic features seen by the XRT in bursts detected by
Swift \cite{zh05a} (see text).}
\end{figure}

Another major advance achieved by Swift was the detection of the long burst 
GRB 050904, which broke through the astrophysically and psychologically 
important redshift barrier of $z\sim 6$. This burst was very bright, 
both in its prompt $\gamma$-ray emission ($E_{\gamma , iso} \sim 10^{54}$ erg) 
and in its X-ray afterglow. Prompt 
ground-based optical/IR upper limits and a J-band detection suggested a 
photometric redshift $z>6$ \cite{haislip05}. Spectroscopic confirmation 
with the 8.2 m Subaru telescope gave a $z=6.29$ \cite{kawai05}.  
There are several striking features to this burst. 
One is the enormous X-ray brightness, exceeding for a full day the X-ray 
brightness of the most distant X-ray quasar know to-date, SDSS J0130+0524, 
by up to a factor $10^5$ in the first minutes \cite{watson05}. The implications 
as a tool for probing the IGM are thought-provoking. Another 
feature is the extremely variable X-ray light curve, showing many large 
amplitude flares extending up to at least a day. A third exciting feature 
is the report of a brief, very bright IR flash \cite{boer05}, comparable in 
brightness to the famous $m_V\sim 9$ optical flash in GRB 990123.

The third major advance from Swift was the discovery and localization of short 
GRB afterglows. As of December 2005 nine short bursts had been localized by 
Swift, while in the same period HETE-2 discovered two, and one was identified 
with the IPN network. In five of the Swift short bursts an X-ray afterglow was 
measured and followed up, with GRB 050709, 050724 and 051221a showing an optical 
afterglow, and 050724 also a radio afterglow, while 040924 had an optical 
afterglow but not an X-ray one \cite{fox05_thisproc}.  These are the 
first afterglows detected for short bursts. Also, for the first time, host 
galaxies were identified for these short bursts, which in four cases are 
early type (ellipticals) and in two cases are irregular galaxies. The redshifts 
of four of them are in the range $z\sim 0.15-0.5$, while another one was 
initially given as $z=0.8$ but more recently has been reported as $z \simeq 1.8$ 
\cite{berger05_thisproc}. The median $z$ is $\siml 1/3-1/2$ that of the long bursts.
There is no evidence for significant star formation in any of these host 
environments, which corresponds to what one would expect for neutron star mergers 
or neutron star-black hole mergers, the most often discussed progenitor candidates 
(it would also be compatible with other progenitors involving old compact stars).

The first short burst seen by Swift, GRB 05059b, was a low luminosity 
($E_{iso}\sim 2\times 10^{48}$ erg) burst with a simple power-law X-ray 
afterglow  which could only be followed for $\sim 10^4$ s \cite{gehr05_0509}.
The third one, GRB 050724, was brighter, $E_{iso}\sim 3\times 10^{50}$ erg, 
and could be followed in X-rays for at least $10^5$ s \cite{barth05_0724}. 
The remarkable thing about this burst's X-ray afterglow is that it resembles 
the typical X-ray light curves described above for long GRB -- except for 
the lack of a slow-decay phase, and for the short prompt emission which 
places it the the category of short bursts, as well as the elliptical host 
galaxy candidate. It also has X-ray flares, at 100 s and another one at 
$3\times 10^4$ s. The first flare has the same fluence as the prompt 
emission, while the late flare has $\sim 10\%$ of that. The interpretation 
of these pose interesting challenges, as discussed below.

\section{Models of early afterglows in light of Swift}
\label{sec:models}

The afterglow is expected to become important after a time
\beq
t_{ag}= {\rm Max}[ (3/4) (r_{dec}/2c \Gamma^2)(1+z)~,~T] =
{\rm Max}[ 10^2 (E_{52}/n_0)^{1/3} \Gamma_2^{-8/3}(1+z)|{\rm s}~,~T]~,
\label{eq:tag}
\enq
where the deceleration time is $t_{dec}\sim (3/4) (r_{dec}/2c \Gamma^2)$ and 
$T$ is the duration of the prompt outflow, $t_{ag}$ marking the beginning of 
the self-similar blast wave regime.

Denoting the frequency and time dependence  of the afterglow spectral energy 
flux as $F_\nu(t)\propto \nu^{-\beta}t^{-\alpha}$, the late X-ray afterglow 
phases (3) and (4)  described above are similar to those known previously 
from Beppo-SAX. (For a review of this earlier behavior and its modeling 
see e.g. \cite{zm04}). The ``normal" decay phase (3), with temporal decay 
indices $\alpha \sim 1.1-1.5$ and spectral energy indices $\beta\sim 
0.7-1.0$, is what is expected from the evolution of the forward shock in the 
Blandford-McKee self-similar late time regime, under the assumption of 
synchrotron emission. 

The late steep decay decay phase (4) of \S \ref{sec:obs}, 
occasionally seen in Swift bursts, is naturally explained as a jet
break, when the decrease of the ejecta Lorentz factor leads to the 
light-cone angle becoming larger than the jet angular extent,
$\Gamma_j(t) \simg 1/\theta_j$ (e.g. \cite{zm04}). It is noteworthy, 
however, that this final steepening has been seen in less than $\sim 10\%$ 
of the Swift afterglows, and then with reasonable confidence mainly in 
X-rays. The corresponding optical light curve breaks have been few, and not 
well constrained.  This is unlike the case with the $\sim 20$ Beppo-SAX 
bursts, for which an achromatic break was reported in the optical 
\cite{frail01}, while in some of the rare cases where an X-ray or radio 
break was reported it occurred at a different time \cite{berger03}. The 
relative paucity of optical breaks in Swift afterglows may be an observational 
selection effect due to the larger median redshift, and hence fainter and
redder optical afterglow at the same  observer epoch, as well as perhaps
reluctance to commit large telescope time on more frequently reported bursts 
(an average, roughly, of 2/month with Beppo-SAX versus 2/week with Swift).

\subsection{Steep decay}
\label{sec:steep}

Among the new early afterglow features  detected by Swift, the steep initial 
decay phase $F_\nu \propto t^{-3}- t^{-5}$ in X-rays of the long GRB 
afterglows is one of the most puzzling. There could be several possible 
reasons for this. The most immediate of these would
be the cooling following cessation of the prompt emission (internal shocks
or dissipation). If the comoving magnetic field in the emission region is
random [or transverse], the flux per unit frequency along the line of sight
in a given energy band, as a function of the electron energy index $p$,
decays as $F_\nu \propto t^{-\alpha}$ with $\alpha={-2p}~[(1-3p)/2]$ in the
slow cooling regime, where $\beta=(p-1)/2$, and it decays as
$\alpha=-2(1+p),~[-(2-3p)/2]$ in the fast cooling regime where $\beta=p/2$,
i.e. for the standard $p=2.5$ this would be $\alpha=-5,~[-3.25]$ in the
slow cooling or $\alpha=-7,~[-2.75]$ in the fast cooling regime, for
random [transverse] fields \cite{mr99}. In some bursts this may be the 
explanation, but in others the time and spectral indices do not correspond well.

Currently the most widely considered explanation for the fast decay, either 
in the initial phase (1) or in the steep flares, attributes it to the 
off-axis emission from regions at $\theta >\Gamma^{-1}$ (the curvature effect,
or high latitude  emission \cite{km00}. In this case, after
the line of sight gamma-rays have ceased, the off-axis emission observed
from $\theta>\Gamma^{-1}$ is $(\Gamma\theta)^{-6}$ smaller than that from
the line of sight. Integrating over the equal arrival time region, this
flux ratio becomes $\propto (\Gamma\theta)^{-4}$. Since the emission from
$\theta$ arrives $(\Gamma\theta)^2$ later than from $\theta=0$, the observer
sees the flux falling as $F_\nu\propto t^{-2}$, if the flux were frequency 
independent. For a source-frame flux $\propto \nu'^{-\beta}$, the observed 
flux per unit frequency varies then as 
\beq 
F_\nu\propto (t-t_0)^{-2-\beta}
\label{eq:hilat}
\enq
i.e. $\alpha=2+\beta$. This ``high latitude" radiation, which for observers 
outside the line cone at $\theta > \Gamma^{-1}$ would appear as prompt 
$\gamma$-ray emission from dissipation at radius $r$, appears to observers 
along the line of sight (inside the light cone) to arrive delayed by $t\sim 
r\theta^2/2c)$ relative to the trigger time, and its spectrum is softened 
by the Doppler factor $\propto t^{-1}$ into the X-ray observer band. 
For the initial prompt decay, the onset of the afterglow (e.g. phases
2 or 3), which also come from the line of sight, may overlap in time
with the delayed high latitude emission.  In equation (\ref{eq:hilat}) 
$t_0$ can be taken as the trigger time, or some value comparable or
less than by equation (\ref{eq:tag}). This can be used to constrain the
prompt emission radius \cite{lazbeg05}. When $t_{dec}<T$, the emission can 
have an admixture of high latitude and afterglow, and since the afterglow 
has a steeper spectrum than the high latitude (which has a prompt spectrum), 
one can have steeper decays \cite{obrien05}. Values of $t_0$ closer to the
onset of the decay also lead to steeper slopes. Structured jets, when
viewed on-beam produce essentially the same slopes as homogeneous jets,
while off-beam observing can lead to shallower slopes \cite{dyks05}.
For the flares, if their origin is assumed to be internal (e.g. some form of
late internal shock or dissipation) the value of $t_0$ is just before the 
flare, e.g the observer time at which the internal dissipation starts to be 
observable \cite{zh05_thisproc}.  This interpretation appears, so far, compatible 
with most of the Swift afterglows \cite{zh05a,nousek05,pan-ag05}. 

Alternatively, the initial fast decay could be due to the emission of 
a cocoon of exhaust gas \cite{peer05}, where the temporal and spectral 
index are explained through an approximately power-law behavior of escape 
times and spectral modification of multiply scattered photons.
The fast decay may also be due to the reverse shock emission, if
inverse Compton up-scatters primarily synchrotron optical photons into
the X-ray range. The decay starts after the reverse shock has crossed
the ejecta and electrons are no longer accelerated, and may have both a
line of sight and an off-axis component \cite{koba05}.
This poses strong constraints on the Compton-y parameter, and cannot
explain decays much steeper than $\alpha=-2$, or $-2-\beta$ if the 
off-axis contribution dominates. Models involving bullets, whose origin, 
acceleration and survivability is unexplained, could give a prompt decay 
index $\alpha=-3-5$ \cite{dado05}, but imply a bremsstrahlung energy index 
$\beta \sim 0$ which is not observed in the fast decay, and require 
fine-tuning. Finally, a patchy shell model, where the Lorentz 
factor is highly variable in angle, would produce emission with 
$\alpha\sim -2.5$. Thus, such mechanisms may explain the more gradual 
decays, but not the more extreme $\alpha=-5,-7$ values encountered in some cases.

\subsection{Shallow decay}
\label{sec:shallow}

The slow decay portion of the X-ray light curves ($\alpha\sim -0.3-0.7$), 
ubiquitously detected by Swift, is not entirely new, having been
detected in a few cases by BeppoSAX. This, as well as the appearance
of wiggles and flares in the X-ray light curves after several hours were
the motivation for the ``refreshed shock" scenario \cite{rm98,sm00}. Refreshed 
shocks can flatten the afterglow light curve for hours or days, even if the 
ejecta is all emitted promptly at $t=T \siml t_\gamma$, but with a range of 
Lorentz factors, say $M(\Gamma) \propto \Gamma^{-s}$, where the lower 
$\Gamma$ shells arrive much later to the foremost fast shells which have already
been decelerated.  Thus, for
an external medium of density $\rho\propto r^{-g}$ and a prompt injection
where the Lorentz factor spread relative to ejecta mass and energy  is
$M(\Gamma)\propto \Gamma^{-s}$, $E(\Gamma)\propto \Gamma^{-s+1}$, the 
forward shock flux  temporal decay is given by \cite{sm00}
\beq
\alpha=[(g-4)(1+s)+\beta(24 -7g +sg)]/[2(7+s-2g)]~.
\label{eq:shallow}
\enq
It needs to be emphasized that in this model all the ejection can be prompt
(e.g. over the duration $\sim T$ of the gamma ray emission) but the low $\Gamma$
portions arrive at (and refresh) the forward shock at late times, which can
range from hours to days. I.e., it is not the central engine which is active
late, but its effects are seen late.  Fits of such refreshed shocks to observed 
shallow decay phases in Swift bursts \cite{grankum05} lead to a $\Gamma$ 
distribution which is a broken power law, extending above and below a peak 
around $\sim 45$.

Another version of refreshed shocks, on the other hand, does envisage central 
engine activity extending for long periods of time, e.g. $\siml$ day (in contrast 
to the $\siml$ minutes engine activity in the model above). Such long-lived 
activity may be due to continued fall-back into the central black hole 
\cite{woo05} or a magnetar wind \cite{zhm01}. 
One characteristic of both types of refreshed models is that after the 
refreshed shocks stop and the usual decay resumes, the flux level shows a 
step-up relative to the previous level, since new energy has been injected.

From current analyses, the refreshed shock model is generally able to explain 
the flatter temporal X-ray slopes seen by Swift, both when it is seen to join 
smoothly on the prompt emission (i.e.  without an initial steep decay phase) 
or when seen after an initial steep decay. Questions remain concerning the
interpretation of the fluence ratio in the shallow X-ray
afterglow and the prompt gamma-ray emission, which can reach $\siml 1$
\cite{obrien05}. This requires a higher radiative efficiency in the prompt
gamma-ray emission than in the X-ray afterglow. One might speculate that this 
might be achieved if the prompt outflow is Poynting-dominated. 
Alternatively, a more efficient afterglow might emit more of its energy in 
other bands, e.g. in GeV, or IR. Or \cite{ioka05} a previous mass ejection
might have emptied a cavity into which the ejecta moves, leading to greater 
efficiency at later times, or otherwise the energy fraction going into the 
electrons increases $\propto t^{1/2}$.

\subsection{X-ray flares}
\label{sec:flares}

Refreshed shocks can also explain some of the X-ray flares whose rise and decay
slopes are not too steep. However, this model encounters difficulties with
the very steep flares with rise or decay indices  $\alpha\sim \pm 5-7$, such
as inferred from the giant flare of GRB 0500502b \cite{burr05a} around 300 s 
after the trigger. Also, the flux level increase in this flare is a factor 
$\sim 500$ above the smooth afterglow before and after it, implying a comparable 
energy excess in the low versus high $\Gamma$ material. An explanation based
on inverse Compton scattering in the reverse shock \cite{koba05} can explain a
single flare at the beginning of the afterglow, with not too steep decay.
For multiple flares, models invoking encountering a lumpy external medium 
have generic difficulties  explaining steep rises and decays \cite{zh05a}, 
although extremely dense, sharp-edged  lumps, if they exist, might satisfy 
the steepness \cite{dermer05}. 

Currently the more widely considered model for the flares ascribes them to 
late central engine activity \cite{zh05a,nousek05,pan-ag05}. 
The strongest argument in favor of this is that the energy budget is 
more easily satisfied, and the fast rise/decay is straightforward to explain.
In such a model the flare energy can be  comparable to the prompt emission, 
the fast rise comes naturally from the short time variability leading to
internal shocks (or to rapid reconnection), while the rapid decay may be
due to the high latitude emission following the flare, with $t_0$ reset to 
the beginning of each flare (see further discussion in \cite{zh05_thisproc}).
However, some flares are well modeled by refreshed forward shocks, while
in others  this is clearly ruled out and a central engine origin is
better suited \cite{wu05}. Aside from the phenomenological desirability 
based on energetics and timescales, a central engine origin is conceivable, 
within certain time ranges, based on numerical models of the core collapse  origin in 
long bursts. These are interpreted as being due to core collapse of a massive 
stellar progenitor, where continued infall into fast rotating cores can continue
for a long time \cite{woo05}. However, large flares with a fluence which 
is a sizable fraction of the prompt emission occurring hours later remain 
difficult to understand. It has been argued that gravitational instabilities 
in the infalling debris torus can lead to lumpy accretion \cite{perna05}.
Alternatively, if the accreting debris torus is dominated by MHD effects,
magnetic instabilities can lead to extended, highly time variable accretion
\cite{proga03}.

\subsection{Short burst afterglows}
\label{sec:short}

Swift, and in smaller numbers HETE-2, have provided the first bona fide 
short burst X-ray afterglows followed up starting
$\sim 100$ s after the trigger, leading to localizations and  redshifts.
In the first of these, GRB 050509b \cite{gehr05_0509} the extrapolation 
of the prompt BAT emission into the X-ray range, and the XRT light curve 
from 100 s to about 1000 s (after which only upper limits exist, even with
Chandra, due to the faintness of the burst) can be fitted with a single
power law of $\alpha \sim 1.2$, or separately as $\alpha_{BAT}=-1.3$
and $\alpha_{XRT}=1.1$. The X-ray coverage was sparse due to orbital
constraints, the number of X-ray photons being small, and no optical
transient was identified, probably due to the faintness of the source.
An optical host was however identified \cite{fox05_0709}, an irregular 
galaxy at $z=0.16$ (and the observations ruled out any supernova association).
On the other hand, GRB 050724 was relatively bright, and besides X-rays, it
also yielded both a decaying optical and a radio afterglow \cite{berg05_4sho}.
This burst, together with the greater part of other short bursts, is
associated with an elliptical host galaxy. It also had a low-luminosity 
soft gamma-ray extension of the short hard gamma-ray component (which would 
have been missed by BATSE), and it had an interesting  X-ray afterglow 
extending beyond $10^5$ s \cite{barth05_0724}. The soft gamma-ray extension, 
lasting up to 200 s, when extrapolated to the X-ray range overlaps well with 
the beginning of the XRT afterglow, which between 100 and 300 s has 
$\alpha\sim -2$, followed by a much steeper drop $\alpha\sim -5-7$ out to 
$\sim 600 s$, then a more moderate decay $\alpha \sim -1$. An unexpected 
feature is a strong flare peaking at $5\times 10^4$ s, whose energy is 
10\% of the prompt emission, while its amplitude is a 10 times increase 
over the preceding slow decay. With about a half dozen of reasonably 
identified objects, the distribution of shorts bursts in redshift space 
and among host galaxy types, including fewer spiral/irregulars and more 
ellipticals, is typical of old population progenitors, such as neutron 
star binaries or black hole-neutron star binaries \cite{nakar05_sho}.

The main challenges posed by the short burst afterglows are the relatively
long, soft tail of the prompt emission, and the strength and late occurrence 
of the flares. A possible explanation for the extended long soft tails ($\sim
100 s$)  may be that the compact binary progenitor is a black hole - neutron
star system \cite{barth05_0724}, for which analytical and numerical arguments
(\cite{davies04}, and references therein)  suggest that the disruption and
swallowing by the black hole may lead to a complex and more extended 
accretion rate than for double neutron stars. The flares, for which the 
simplest interpretation might be as refreshed shocks (which would be 
compatible with a short engine duration $T\siml t_\gamma \sim 2$ s, for a 
Lorentz factor distribution), requires the energy in the slow material to 
be at least ten times as energetic as the fast material responsible for the 
prompt emission, for the GRB 050724 flare at $10^4$ s. The rise and decay 
times are moderate enough for this interpretation. Another interpretation
might be an accretion-induced collapse of a white dwarf in a binary, leading
to a flare when the fireball created by the collapse hits the companion 
\cite{macfad05}, which might explain moderate energy one-time flares.
However, for repeated, energetic flares, as also in the long bursts, the 
total energetics are easier to satisfy if one postulates late central 
engine activity (lasting at least half a day), containing $\sim 10\%$ of 
the prompt fluence \cite{barth05_0724}. A possible way to produce this 
might be temporary choking up of an MHD outflow \cite{proga03} (c.f. 
\cite{vanput01}), which might also imply a linear polarization of the 
X-ray flare \cite{fanproga05}. Such MHD effects could plausibly also explain 
the initial $\sim 100$ s soft tail. However, a justification for substantial
$\simg 10^5$ s features remains so far on tentative grounds.

The similarity of the X-ray afterglow light curve with those of long bursts 
is, in itself, an argument in favor of the prevalent view that the afterglows 
of both long and short bursts can be described by the same paradigm, 
independently of any difference in the progenitors. This impression is 
reinforced by the fact that the X-ray light curve temporal slope is, on 
average, that expected from the usual forward shock afterglow model, and that 
in two short bursts (so far) there is evidence for what appears to be a jet 
break \cite{berg05_4sho,pan05_sho}. However, while similar to zeroth order,
the first order differences are revealing: the average isotropic energy is 
factor $\sim 100$ smaller, while the average jet opening angle (based
on two breaks) is a factor $\sim 2$ larger \cite{fox05_thisproc,pan05_sho}.
Using the standard afterglow theory, the bulk Lorentz factor decay can be 
expressed through $\Gamma(t_d) =6.5(n_o/E_{50})^{1/8} t_d^{-3/8}$, where 
$t_d=(t/{\rm day})$, $n_o$ is the external density in units of cm$^{-3}$,
and $E_{50}$ is the isotropic equivalent energy in units of $10^{50}$ ergs.
If the jet break occurs at $\Gamma(t_{br})=\theta_j^{-1}$ the jet opening 
angle and the total jet energy $E_j$ are
\beq
\theta_j=  9^o (n_o/E_{50})^{1/8} t_{d,br}^{3/8}~~,
E_j =  \pi \theta_j^2 E \sim 10^{49} n_o^{1/4} (E_{50} t_{d,br})^{3/4}~{\rm erg}~.
\label{eq:short}
\enq
For the first two well studied afterglows GRB 050709 and GRB 050724, 
together with the standard afterglow expressions for the flux level as
a function of time before and after the break, this leads to fits 
\cite{pan05_sho} which are not completely determined, allowing for GRB050709
either a very low or a moderately low external density, and for GRB050724 a 
moderately low to large external density. The main uncertainty is in the
jet break time, which is poorly sampled, and so far mainly in X-rays.
As brighter short bursts are occasionally detected, e.g. GRB 051221A, the
chances of tighter constraints on jet breaks should increase.

\subsection{Prompt optical flashes and high redshift afterglows}
\label{sec:promptopt}

Optical/UV afterglows have been detected with the Swift UVOT telescope in
roughly half the bursts for which an X-ray afterglow was seen. For a more
detailed discussion of the UVOT afterglow observations see \cite{roming05}.
Of particular interest is the ongoing discussion on whether ``dark GRB" are 
really optically deficient, or the result of observational bias \cite{berger05a}.
Another puzzle is the report of a bimodal intrinsic brightness distribution in 
the rest-frame R-band \cite{liangzh05,nard05}.  This suggests possibly the 
existence of two different classes of long bursts, or at least two different 
types of environments. 

Compared to a few years ago, a much larger role is being played by 
ground-based robotic optical follow-ups, due to the increased rate of 
several arc-second X-ray alerts from XRT, and the larger number of robotic 
telescopes brought on-line in the last years. For the
most part, these detections have yielded optical decays in the $\simg$ few
100 s range, initial brightness $m_V\sim 14-17$ and temporal decay slopes 
$\alpha\sim 1.1-1.7$ previously associated with the evolution of a forward 
shock \cite{fox05_thisproc,berger05_thisproc}, while in a few cases a 
prompt optical detection was achieved in the first 12-25 s.

The most exciting prompt robotic IR detection (and optical non-detection)
is that of GRB 050904 \cite{boer05,haislip05}. This object, at the
unprecedented high redshift of $z=6.29$ \cite{kawai05}, has an X-ray brightness 
exceeding for a day that of the brightest X-ray quasars \cite{watson05}, and 
its O/IR brightness in the first 500 s (observer time) was comparable to that
of the extremely bright ($m_V\sim 9$) optical flash in GRB 990123, with a 
similarly steep time-decay slope $\alpha\sim 3$ \cite{boer05}. Such prompt, 
bright and steeply decaying optical emission is expected from the reverse shock 
as it crosses the ejecta, marking the start of the afterglow \cite{mr97a}. 
However, aside from the two glaring examples of 990123 and 05094, in the last
six years there have been less than a score of other prompt optical flashes, 
typically with more modest initial brightnesses $m_v\simg 13$. There are a  
number of possible reasons for this paucity of optically bright flashes, if
ascribed to reverse shock emission. One is the absence or weakness of a reverse 
shock, e.g. if the ejecta is highly magnetized \cite{mr97a}. A moderately 
magnetized ejecta is in fact favored for some  prompt flashes \cite{zkm03}. 
Alternatively, the deceleration might occur in the thick-shell regime 
($T \gg t_{dec}$. see eq. (\ref{eq:tag}), which can result in the reverse 
shock being relativistic, boosting the optical reverse shock spectrum into 
the UV \cite{k00}. Another possibility, for a high comoving luminosity, 
is copious pair formation in the ejecta, causing the reverse shock spectrum 
to peak in the IR \cite{mrrz02}. Since both GRB 990123 and GRB 050904 had 
$E_{iso}\sim 10^{54}$ erg, among the top few percent of all bursts, the latter 
is a distinct possibility, compatible with the fact that the prompt flash in 
GRB 050904 was bright in the IR I-band but not in the optical. On the other 
hand, the redshift $z=6.29$ of this burst, and a Ly-$\alpha$ cutoff at 
$\sim 800$ nm would also ensure this (and GRB 990123, at $z=1.6$, was detected 
in the V-band). However, the observations in these two objects but a suppression 
in lower $E_{iso}$ objects appears compatible with having a relativistic 
(thick shell) reverse shock with pair formation.

\begin{theacknowledgments}
I am grateful to the Swift team for collaborations and to 
NASA NAG5 13286 for support. 
\end{theacknowledgments}

\bibliographystyle{aipproc}   

\end{document}